\documentclass[a4paper,11pt]{article}
\usepackage{jcappub} % for details on the use of the package, please see the JINST-author-manual
\usepackage{lineno}
\usepackage[T1]{fontenc} % if needed
\usepackage{subfigure}

\usepackage{enumitem,amssymb,amsmath}

\arxivnumber{2401.14431} % Only if you have one

\title{\boldmath Primordial naked singularities}

\author[a]{Pankaj S. Joshi}
\author[b]{and Sudip Bhattacharyya}
\affiliation[a]{International Centre for Space and Cosmology, School of Arts and Sciences, Ahmedabad University, Ahmedabad 380009, India}
\affiliation[b]{Department of Astronomy and Astrophysics, Tata Institute of Fundamental Research, 1 Homi Bhabha Road, Colaba, Mumbai 400005, India}
\emailAdd{pankaj.joshi@ahduni.edu.in}
\emailAdd{sudip@tifr.res.in}

\abstract{Primordial black hole formation has been discussed widely, when density perturbations in the early universe cause matter to collapse gravitationally, giving rise to these ultra-compact objects. We propose and point out that such a gravitational collapse would also give rise to primordial naked singularities, that would play an important role in the observable features of present universe. We consider two types of collapse scenarios that give rise to event-like and object-like visible singularities within a cosmological background. We briefly discuss implications of primordial naked singularities, including those for dark matter, vis-a-vis primordial black holes.}


\begin{document}
\maketitle
\flushbottom

\section{Introduction}\label{Introduction}

Gravitational collapse is a fundamental process in the universe, playing basic role for formations of large scale structure, galaxies, and clusters of galaxies.   At more local scales stars also form due to collapse of matter 
clouds in star-forming regions. Again, such collapse governs the final fate of massive stars when they exhaust their internal nuclear fuel.  
Moreover, density perturbations in the early universe cause gravitational collapse in local regions, which is believed to have given rise to primordial black holes (PBHs) \cite{ZeldovitchNovikov1966,Hawking1971}. 
Such PBHs could have formed in the very early universe, perhaps less than one second after the big bang, during the inflationary era, or in very early radiation-dominated era.  
A typical density contrast of ${\displaystyle \delta \rho /\rho \sim 0.1}$ ($\rho$ being the average density of the universe) induces  gravitational collapse to form a PBH. The physical implications of these ultra-compact objects are widely discussed and have attracted much attention \cite{Villanueva-Domingo2021,Kokubu2018,Khlopov2010,Belotskyetal2014,Belotskyetal2019}. 

Towards the end of its life cycle, catastrophic gravitational collapse of a massive star takes place. Under physically reasonable conditions,  such a collapse would always result in a space-time singularity, as shown by the singularity theorems in general relativity. Just as in gravitational collapse, such spacetime singularities would occur in cosmology as well \cite{Hawking1973}. 
An important point here is, while the singularity theorems predict the occurrence of spacetime singularities, they give no information on the nature or causal structure of the singularities, and in particular, whether these could be visible to the faraway observers in the universe. It is often believed or conjectured that gravitational collapse of massive stars necessarily ends in black hole formation only, which is the scenario when singularities are always covered by the event horizons of gravity.

In order to understand the final fate of such a gravitational collapse, we need to probe and work out the dynamics and evolution of collapse processes in general relativity. Such an effort has taken place in past years and the collapse dynamics has been examined in detail. The outcome is that such a singularity would be hidden within an event horizon of gravity, or could occur outside the event horizon, thus being a visible or naked singularity. 
When covered within an event horizon, the singularity is not seen by faraway observers. However, when not hidden within the horizon, such a visible singularity could send observable signals from the ultra-strong gravity regions to faraway observers in the universe. In fact, if the collapsing cloud is not homogeneous, the collapsing matter shells could cross each other, thus causing the so called `shell-crossing' naked singularities 
\cite{Yodzisetal1973}.
Such shell-crosses could occur at various places in the spacetime depending on the inhomogeneous matter distribution. As opposed to this, naked singularities at the centers of inhomogeneous, collapsing dust clouds were shown to be occurring as well, which were considered to be more serious \cite{Christodoulou1984}. 
This class of naked singularities was also reported earlier using numerical methods \cite{EardleySmarr1979}.
However, as it turned out, each of these classes of naked singularities were seen to be `gravitationally weak', in the sense that the spacetime curvatures did not diverge sufficiently strongly in their vicinity \cite{Newman1986a}. 
In that case, the spacetime would be regular enough even near these naked singularities and could then be extended through the same to cure or remove the singularities. These developments gave rise to many related investigations on inhomogeneous collapse (for further details see, e.g., 
\cite{Joshi1993}, and references therein). 
Eventually, a full analysis of the inhomogeneous dust collapse involving the complete parameter space of the mass and velocity functions for this spacetime, known as the Lemaitre-Tolman-Bondi models, was given and the classes of strong curvature naked singularities arising were identified
\cite{JoshiDwivedi1993}, 
which were seen to be genuine naked singularities arising from reasonable and regular physical initial conditions, which were not removable like the earlier classes mentioned above.

Extending these results, many realistic gravitational collapse scenarios such as collapse of perfect fluids with a reasonable equation of state, scalar field collapse, self-similar collapse, non-spherical collapse and other models including numerical simulations of collapse have been investigated and analyzed to examine the final state of a gravitational collapse in terms of occurrence of a black hole or a naked singularity
\cite{HellabyLake1985, GoncalvesJhingan2001, Giamboetal2003, Haradaetal2002, OriPiran1990, Gundlach1999, Christodoulou1999, Nolan2002, Giacomazzoetal2011, Ortiz2011, Banerjeeetal2003, Barausseetal2010}.
It is seen that the final state singularity is visible under physically reasonable conditions. The typical result is that the generic regular initial conditions for the collapsing matter, such as the density and pressure distributions and the initial velocity profiles, from which the collapse develops could give rise to naked singularities just as they could develop into black holes (for further details see, e.g., 
\cite{JoshiMalafarina2011} and references therein).

Within such a perspective, a question that arises naturally is, what are the physical conditions and triggers that give rise to a naked singularity as the final state of collapse, rather than a black hole. In other words, why would a naked singularity develop at all in collapse, rather than a black hole?
To probe this issue we could consider an explicit example. For a homogeneous pressureless cloud which has a constant density from center to the boundary of the cloud, the final singularity will be necessarily hidden within the event horizon 
\cite{Oppenheimer1939, JoshiDwivedi1999}. 
The same result holds for a non-rotating homogeneous perfect fluid collapse or other matter fields as well.
However, this is an idealized situation, stars typically have density higher at their center that decreases towards the surface. It is seen that for gravitational collapse of such an inhomogeneous cloud the outcome is a visible or naked singularity \cite{JoshiDwivedi1999,JoshiDwivedi1993,GoswamiJoshi2007}.

We thus find the theoretical studies made so far implying that the occurrence of naked singularities as final collapse states happens in a wide variety of gravitational collapse scenarios.  
Indeed, some specific collapsed objects have been proposed to be naked singularities. For example, the Event Horizon Telescope
Collaboration have discussed \cite{eht2022} the possibility of Sgr A* being a first type of
Joshi-Malafarina-Narayan (JMN-1; \cite{jmn11}) naked singularity. 
Moreover, the supermassive collapsed object M87* and the stellar-mass collapsed object in the X-ray binary GRO J1655--40 could be naked singularities \cite{gcyl,cbgm}.

We therefore propose here that just as the 
PBHs are hypothetically suggested to have been formed soon after the initial Big Bang singularity, similarly, primordial naked singularities (PNaSs) would form in abundance in the early universe. 
Ultra-high densities and heterogeneous conditions were abound in the early universe causing density fluctuations, so the theoretical collapse models strongly suggest that when sufficiently dense local regions underwent gravitational collapse, they could typically form visible naked singularities. The masses and timescales of such PNaSs forming would be distributed over a wide range, similar to that proposed for PBHs, as the basic physical process forming either of these entities remains the same, namely the general relativistic gravitational collapse.

If PNaSs formed abundantly in the early universe from collapsing density fluctuations, the natural question arising is that of their lifetimes and observational consequences. A question of much interest would be, if PNaSs developed in the early universe, what physical cosmic problems that would help solve. One would like to know how to distinguish a population of these objects vis-a-vis the hypothetical case of only PBHs occurring. While theoretically either can occur under reasonable physical conditions in the early universe, we need to decide which of these entities would be more likely to form.

\section{Gravitational Collapse}\label{Collapse}

As pointed out above, naked singularities are found to occur as gravitational collapse final states in a wide variety of gravitational collapse models. In such a case, an important question that arises is, how generic is the naked singularity formation in collapse, and if they do occur rather generically, if that would break the predictability in the spacetime, which is usually thought to be a typical feature of a classical theory such as the general theory of relativity.  

In fact these issues have been discussed and examined in some detail in the work so far on gravitational collapse in general relativity. 
Whereas a detailed analysis and discussion on these issues would not be relevant here, we mention a few typical results in this context. 
Naked singularities and other features especially for self-similar gravitational collapse in general relativity were analyzed by Ori and Piran \cite{OriPiran1990}, 
who showed for a perfect fluid with a barotropic equation of state
that there is the generic appearance of naked singularities in these solutions.
As they noted, these solutions are quite different from the homogeneous Oppenheimer-Snyder dust collapse model on which our intuition on black holes is so heavily based.
Again, Generic naked singularities in Vaidya spacetimes
are analyzed by
\cite{Wheeler2022}
who showed that globally naked singularities are significantly more common and generic in a natural topology on this collection of spacetimes, and that the curvature strength of the singularity cannot be smoothed away. Again, Zhang \cite{Zhang2017} pointed out that naked singularity naturally emerges from reasonable initial conditions in the collapse processes.
Sadhu and Vardarajan
\cite{Sadhu2013}
showed the stability of Janis-Newman-Winicour naked singularity under scalar field perturbations. 
Also, Joshi, Malafarina and Saraykar \cite{JoshiMalafarinaSaraykar2012}
investigated the genericity and stability aspects for naked singularities and black holes for general Type I matter field collapse that includes most of the physically reasonable matter fields, showing that each one are generic outcomes of collapse, when genericity is defined in a suitable sense in an appropriate topological space.

As for determinism and predictability, it is worth noting that general relativity is in any case not predictable in the usual sense as understood for a classical Newtonian theory. As shown by Reall \cite{Reall2018}, 
numerical analysis of perturbations of a charged black hole suggested that the usual predictability of the laws of physics can fail in general relativity.
Traditional determinism is expressed by the global hyperbolicity condition in general relativity
\cite{Hawking1973},
which is formulated in terms of existence of a global Cauchy surface in the spacetime from which all past and future can be predicted in an evolutionary manner. 
However, as pointed out there, only the simplest of the spacetimes such as Minkowski or Friedmann models are deterministic or predictable in this sense, namely that of admitting a global Cauchy surface. Almost all other models of interest such as the Reissner-Nordstrom, Kerr and others are not deterministic in this sense, the latter in fact admitting closed timelike curves or causality violations near the singularity. In fact, this was the real reason and motivation to create and generalize the singularity theorems by Hawking and Penrose in 1970s, because the first such theorem given by Penrose in 1965 assumed global hyperbolicity or complete determinism, which was thought to be a too restrictive condition on a spacetime model of the universe
\cite{Hawking1973, Joshi1993}.

Thus the generality and stability of naked singularities have been discussed in a number of papers and it is seen that these do occur, satisfying various energy conditions with regular initial data, and that these states constitute a non-zero measure set (see also \cite{Joshi1993,JoshiDwivedi1999,Engelhardtetal}, and references therein). We thus see that detailed analytic results on collapse in four space-time dimensions show generic occurrence of naked singularities, and also numerical simulations by various groups show naked singularity occurrence in higher dimensions as well
(see, e.g. \cite{Yamada2011} and references therein).

We now consider below two scenarios typical of the wide classes of naked singularities that develop in gravitational collapse of different varieties of matter fields.
In order to explore the PNaS formation in the early universe context, we examine here collapse scenarios that give rise to event-like or object-like naked singularities, covering matter fields such as dust, perfect fluids and scalar fields collapse. It turns out that in the theoretical models investigated so far, two main types of naked singularities are found to occur, one that is short-lived or event-like, and others are eternal long-lasting object-like variety. Also, the universe in its very early phase is believed to be dominated by abundance of scalar fields with possible accompanied potentials, hence their collapse would be relevant for PNaSs. The typical scenario here is that of a blob of matter or density fluctuation collapsing locally, within the overall background of a Friedmann-Lemaitre-Robertson-Walker (FLRW) cosmological model.

\subsection{Event-like Singularities}\label{Event}

A remarkable feature of many dust and perfect fluid collapse scenarios is that, when energy conditions ensuring positivity of mass and energy density are satisfied, then singularity becomes visible when it forms. Families of null and timelike geodesics and non-spacelike curves emerge from the same that would reach faraway observers at infinity. At later times the singularity gets covered within the event horizon, and therefore this is an {\it event-like naked singularity}. This renders ultrahigh energy density and curvature regions visible to external observers, which may have important physical implications. 

To get an insight into the formation process of PNaS and to give a typical example, we consider gravitational collapse 
of spherically symmetric inhomogeneous fluid. This helps understand how naturally the PNaS may arise as gravitational collapse final states.
The components of stress-energy tensor in the coordinate basis $\{dx^{\mu} \bigotimes \partial_{\nu} \vert 0 \leq \mu, \nu \leq 3\}$ of the comoving coordinates 
$(t,r,\theta,\phi)$  are given by
$T^{\mu}_{\nu}=\textrm{diag}\left(-\rho,p,p,p\right)$
where $\rho=\rho(t,r)$ and $p=p(t,r)$ are density and isotropic pressure of the collapsing matter cloud respectively. The corresponding spacetime, the Lemaitre-Tolman-Bondi (LTB) metric, is given by,
\begin{equation}
ds^2 = -dt^2 + R^{\prime 2}dr^2 + R^2 d\Omega ^2, \label{spacetime}
\end{equation}
where $d\Omega^2$ is line element of two-sphere at $R=R(t,r)$. 
Then the density and pressures are given by
\begin{equation}
\rho = \frac{F^{'}}{ R^{'} R^2},\\\
     p= - \frac{\dot{F}}{\dot{R} R^2},
\end{equation}
where $F(r,t) = \dot{R}^2 R$ is the Misner-Sharp mass  
of the collapsing cloud giving total mass within a coordinate radius $r$,  
the dot and prime denote partial derivatives with respect to time $t$ and 
radial coordinate $r$.  The collapsing cloud consists of concentric spherical shells, each identified by a radial coordinate $r$. 

For the pressureless marginally bound dust case, $p=0$, and we get $F=F(r)$ 
\cite{JoshiDwivedi1999,JoshiDwivedi1993,GoswamiJoshi2007}, 
and one obtains
\begin{equation}
R(t,r)  = \left( r^\frac{3}{2} - \frac{3}{2} \sqrt{F(r)}t\right)^\frac{2}{3}. \label{physicalr}
\end{equation}
%Defining a \textit{scaling function} $a(t,r)$ as $a(t,r)=\frac{R(t,r)}{r}$, 
%one can rewrite Eq. (\ref{physicalr}) to obtain the time curve,
%\begin{equation}\label{timecurve}
 %   t(r,a)=\frac{2r^{3/2}}{3\sqrt{F(r)}}\left(1-a^{3/2}\right).
%\end{equation}

Given a spherical shell of fixed radial coordinates, as one evolves the initial data, the physical radius of the shell decreases and becomes zero at the singularity. The corresponding comoving time $t_s(r)$ is obtained by substituting $R(t_s,r)=0$ in Eq. (\ref{physicalr}) as,
\begin{equation}
t_{s}(r) =  \frac{2r^{\frac{3}{2}}}{3\sqrt{F(r)}}. 
\end{equation}
This gives the \textit{singularity curve} depicting the singularity
forming as collapse endstate. The singularity time is then no longer a constant here as in the case of the Oppenheimer-Snyder homogeneous dust collapse model but varies with $r$, because concentric shells of matter at different radii arrive one after the other at the central singularity at $r=0$.

In order to model the PNaS formation, a piecewise LTB cell-model for the universe, incorporating local collapsing inhomogeneities, made up of LTB underdense and overdense regions, surrounded by an intermediate region of LTB shells embedded in an expanding FLRW universe 
can be used \cite{Chamorro2001}. 
This generalizes the Einstein-Straus swiss-cheese models, and incorporates the time evolution of cosmological inhomogeneities, thus allowing for collapse of a massive dust cloud in such a cosmological framework. It is then seen that the collapsing local inhomogeneities in an expanding universe could result in either a black hole or a naked singularity, depending on the nature of initial data, which consists of density distribution and the velocities of collapsing shells, from which the collapse commences.

To understand the process of naked singularity formation as collapse final state in the above scenario, we note that the key entity describing the causal structure of such a dynamical collapse is the {\it apparent horizon} that forms as the collapse develops, where the expansion of the outgoing null geodesic congruence vanishes. 
As the collapse initial data evolves from an initial spacelike surface, the evolution of apparent horizon is obtained from (\ref{physicalr}), by equating $F=R$, to get, 

%%%%%%%%%%%%%%%%%%%%%%%%%%%%%%%%%%%%%%%%%%%%%%%%%%%%%%%%%%%%%%%%%%%%%%%%%%%%%%%%%
\begin{figure}[htbp]
\centering
\includegraphics[width=0.7\textwidth]{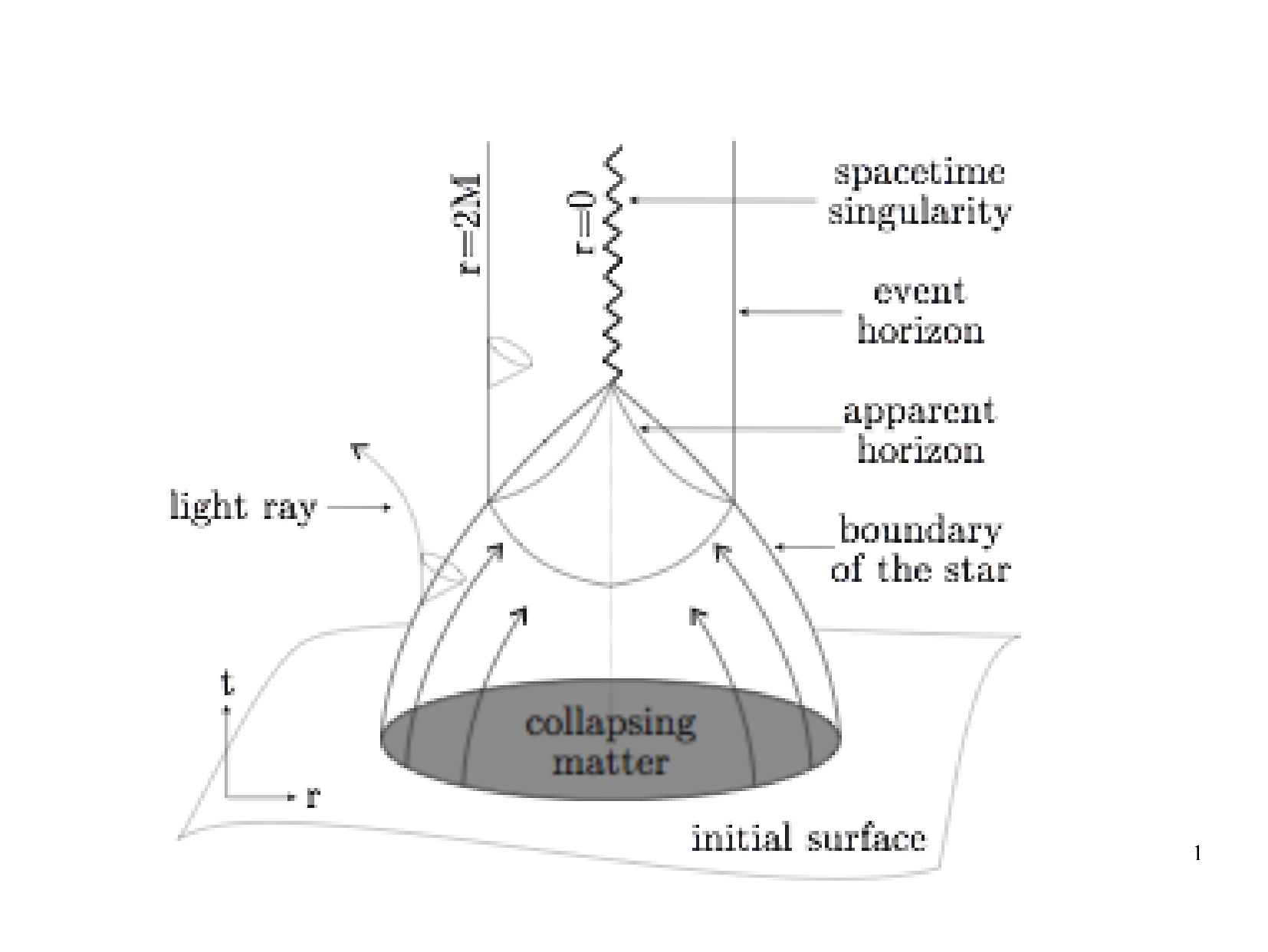}
\caption{Spatially homogeneous collapse where outcome is a black hole.\label{HomoLTB}}
\end{figure}
%%%%%%%%%%%%%%%%%%%%%%%%%%%%%%%%%%%%%%%%%%%%%%%%%%%%%%%%%%%%%%%%%%%%%%%%%%%%%%%%%

\begin{equation}\label{ltbtimeah}
    t_{ah}(r) =\frac{2}{3}\left(\frac{r^{3/2}}{\sqrt{F}}-F\right).
\end{equation}
This \textit{apparent horizon curve} gives the relation between the radial coordinate and comoving time $t$, which actually gives the boundary of trapped surfaces forming in the spacetime as the collapse progresses.
The collapsing fluid spacetime Eq. (\ref{spacetime}) can be matched smoothly with the FLRW model as described above, so that their union forms a full solution of Einstein equations, describing local density inhomogeneities collapsing in a cosmological background. 

%%%%%%%%%%%%%%%%%%%%%%%%%%%%%%%%%%%%%%%%%%%%%%%%%%%%%%%%%%%%%%%%%%%%%%%%%%%%%%%
\begin{figure}[htbp]
\centering
\includegraphics[width=0.7\textwidth]{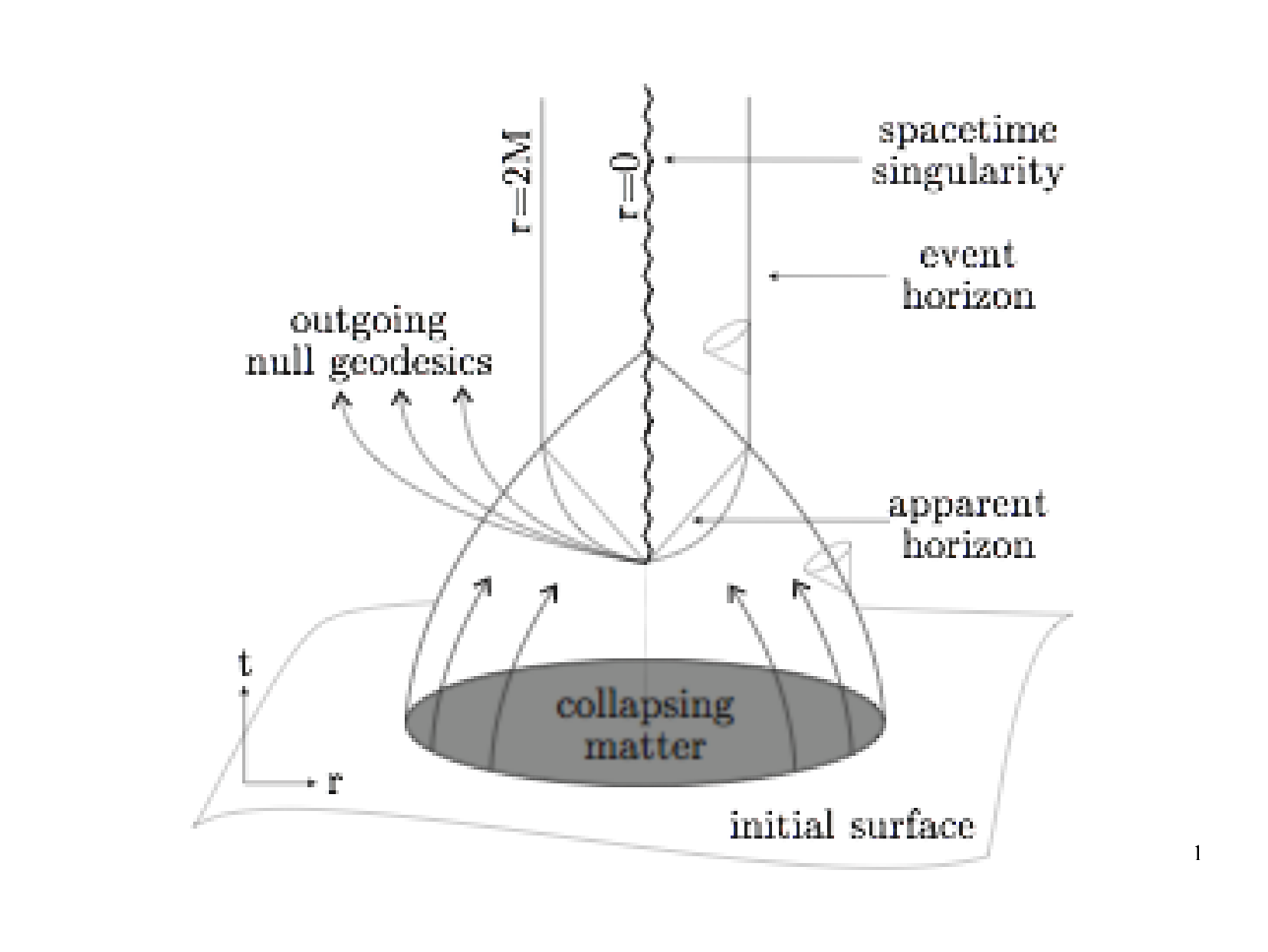}
\caption{Spatially inhomogeneous collapse where outcome is a naked singularity.\label{InHomo}}
\end{figure}
%%%%%%%%%%%%%%%%%%%%%%%%%%%%%%%%%%%%%%%%%%%%%%%%%%%%%%%%%%%%%%%%%%%%%%%%%%%%%%%

In order to understand the formation dynamics of how the singularity occurs first and
then the event horizon appears later that covers the singularity, let us consider the following for the collapsing inhomogeneous dust cloud above. In order to consider physically reasonable configurations let us take the 
density for the collapsing cloud higher at the center.  A generic (inhomogeneous) mass profile for the cloud 
has the form
\begin{equation}
F(r)= F_0r^3 + F_1 r^4 + F_2 r^5 +\cdots\,,
\end{equation}
near $r=0$, where $F_0=\rho_{\rm c}/3$. Here the homogeneous Oppenheimer-Snyder dust collapse corresponds to $F_n=0$ for $n>0$.  Clearly, if there is a negative density gradient, that is, when there is a higher density at the center, then $F_n\neq0$ for some $n>0$. We may note that if we desire the density profile to be analytic, then all odd terms $F_{2n-1}$ would be vanishing.

The important question now is, what is the effect of such an
inhomogeneity on the evolution and development of the trapped surfaces, or in other words, for the apparent horizon. 
To see this, we can determine the behavior of apparent horizon
in the vicinity of the central singularity at $R=0,r=0$. 
For this, let the first non-vanishing derivative of the density at $r=0$ be the $n$-th one ($n>0$), i.e.,
\begin{equation}
F(r) = F_0 r^3 + F_n r^{n+3}+\cdots\,,~~ F_n<0\,
\end{equation}
near the center. The apparent horizon equation above then becomes,
\begin{equation}
t_{\rm ah}(r) = t_0 - {2 \over 3}F_0 r^3 - {F_n \over
3{F_0}^{3/2}} r^n + O(r^{n+1}),
\end{equation}
where $t_0$ is the time of formation of the central singularity at $r=0$. The homogeneous case is given by $F_n=0$ for all values of $n$. The apparent horizon dynamics can now be understood from the above for the homogeneous as well as the inhomogeneous cases. In both cases, the apparent horizon begins at the central singularity $r=0$ at $t=t_0$. In the homogeneous case, it is decreasing in time away from the center for higher values of the radial coordinate $r$, finally meeting the boundary of the cloud well before the singularity formation time. Since the event horizon must cover all trapped surfaces as well as the apparent horizon, it forms well before the singularity in this homogeneous case, thus hiding the singularity and giving rise to a black hole (Fig.~1). On the other hand, for an inhomogeneous collapse, the apparent horizon is seen to be increasing in time away from the center. For example, for a typical density profile such as $\rho(r) = \rho_0 + \rho_2 r^2$ with $\rho_2<0$, from equation (2.2) we see that $n=2$, thus making the apparent horizon curve increasing in time because $F_2<0$.

In this latter case of inhomogeneous collapse, we see that the apparent horizon meets the boundary of the cloud at an epoch later than the central singularity formation time, where it also meets the event horizon (Fig.~2). In this case, when we examine the outgoing families of the timelike and null geodesics in the spacetime, we find that there are families of infinitely many such curves existing such that in the past they meet the first point of the singularity $t=t_0$, and in future they go away to infinity. All these curves of course must necessarily lie outside the event horizon, else they would not travel to infinity but will be hidden within.
In fact, in this case the event horizon is the last null geodesic to come out of the singularity, which first increases in radial coordinate $r$ as it moves away from the singularity, but finally stays put at $r=2m$ as it reaches the boundary of the 
cloud. These features are depicted in Fig.~2. It follows that there is a whole family of infinitely many timelike and null geodesics that comes out of the singularity before the event horizon emerges, which reach the observer at infinity. This happens because this first point of singularity curve turns out to be a `nodal point' for the first order differential equation that describes the null or timelike geodesics in the spacetime.

%%%%%%%%%%%%%%%%%%%%%%%%%%%%%%%%%%%%%%%%%%%%%%%%%%%%%%%%%%%%%%%%%%%%%%%%%%%%%%%
\begin{figure}[htbp]
\centering
\includegraphics[width=0.8\textwidth]{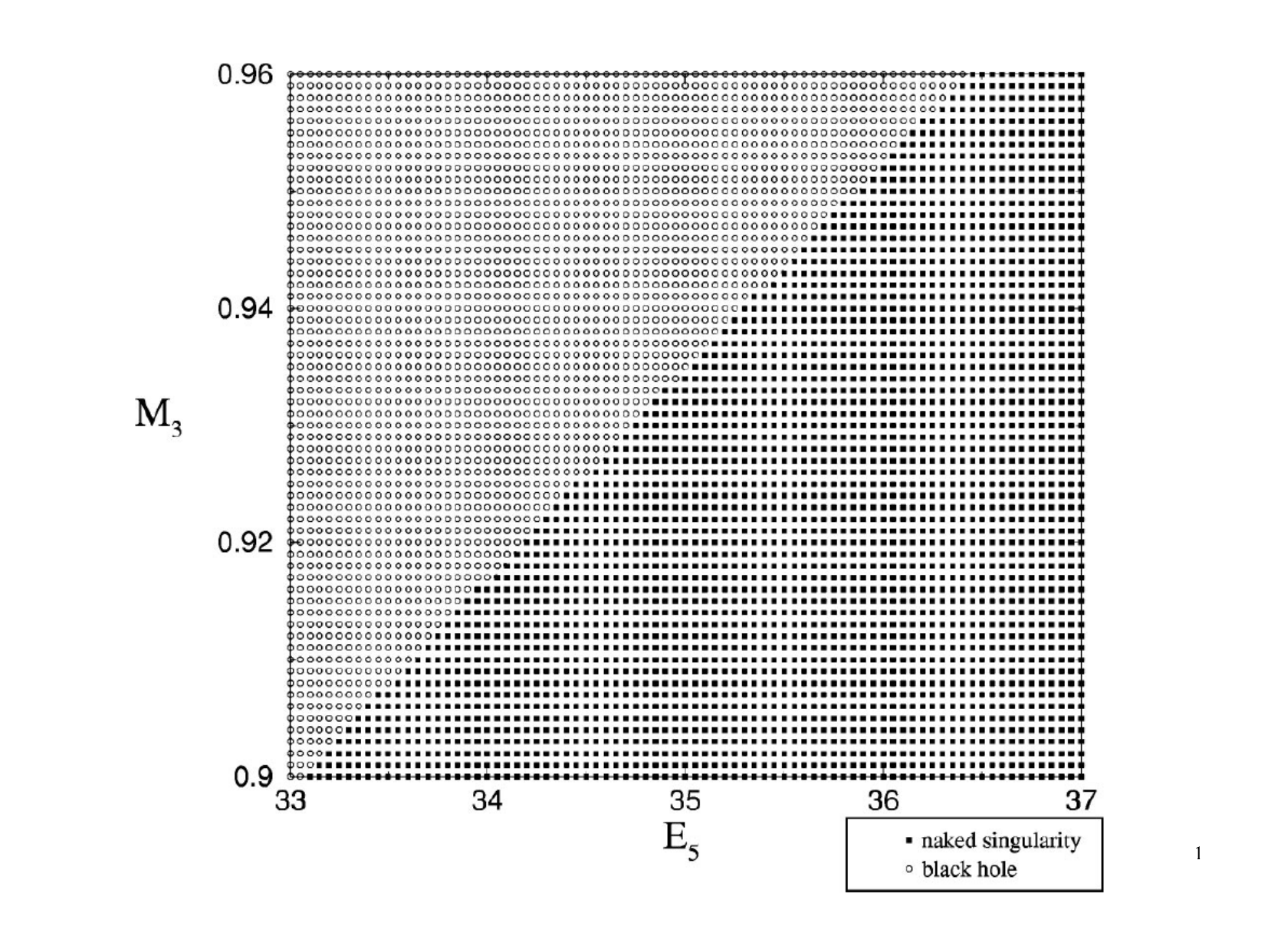}
\caption{Occurrence of black holes and naked singularities as collapse final states for typical density and velocity distributions, indicated by $M$ and $E$, of collapsing shells (from Mena, Tavakol and Joshi \cite{Menaetal2000}).\label{BH-NS}}
\end{figure}
%%%%%%%%%%%%%%%%%%%%%%%%%%%%%%%%%%%%%%%%%%%%%%%%%%%%%%%%%%%%%%%%%%%%%%%%%%%%%%%

We note that Figs. \ref{HomoLTB} and \ref{InHomo} depict spacetime diagrams of evolutions of the apparent horizon and event horizon, when the collapse leads to a black hole and
a globally visible naked singularity, respectively.  
We see that, when the matter distribution is completely homogeneous, the collapse gives rise to a black hole final state, but when the density is higher at the center of the cloud, a physically more realistic case, then naked singularity results. In this latter case the causal structure of collapse and event horizon is so altered to give rise to a visibile singularity. In this situation, the light rays from singularity reach distant observers and the outcome is an event-like naked singularity.

In general, when we consider both the density and velocity profile functions for the collapsing matter shells, 
the distribution of black holes and naked singularities as collapse final states is described by Fig.~3 for typical such profiles
\cite{Menaetal2000}. 
This example also illustrates the genericity of occurrence of naked singularities as gravitational collapse final states in this scenario.

\subsection{Object-like Singularities}\label{Object}

The object-like naked singularities that stay on and continue to exist once created in early universe, could be of more interest from the perspective of cosmological scenarios. Such singularities, if they formed in early universe, will keep evolving with the evolving universe. So their nature and lifetime is very different from the momentary or event-like naked singularities, and they could be eternal or with a nonzero finite lifetime, with various observational consequences to be examined.

Such models are obtained by the JMN gravitational collapse \cite{Joshi2014} 
that creates object-like naked singularities. As the collapse progresses here, the non-zero pressures within the cloud slow down the collapse without allowing for trapped surfaces or apparent horizon to occur. Then the extreme density region that forms at the center continues to be visible eternally. A new procedure was developed here to generate equilibrium configurations that result as the final state of collapse from regular initial conditions. The equilibrium geometries so generated are static solutions with a timelike naked singularity as final state. The collapsing matter could be with non-vanishing tangential pressures, or perfect fluids and others.
  
The formation of these object-like naked singularities is outlined 
below. The spherically symmetric metric for a dynamical collapse can be written as,
\begin{equation}\label{metric}
    ds^2=-e^{2\nu}dt^2+\frac{R'^2}{G}dr^2+R^2d\Omega^2 \; ,
\end{equation}
where $\nu$, $R$ and $G$ are functions of the comoving time $t$ and
radial coordinate $r$.  In the case of
vanishing radial pressure, the energy-momentum tensor is given by
$T^0_0=\rho, \; T_1^1=0, \; T_2^2=T_3^3=p_\theta$. 
To examine the collapse evolution, 
we note there are six unknowns, $\rho$, $p_\theta$, $\nu$, $G$, 
$R$, and the mass function $F$ for the cloud, coming from the integration of the four Einstein equations. We therefore have the freedom to choose 
two free functions.  Once the initial data for the above six functions 
is specified at an initial epoch $t=t_i$ and the two free functions 
specified, the system is closed and Einstein equations 
evolve the collapse to a future time. 
A choice of $F(r)$ and the equilibrium metric function $v_e(r)$ can be made, that closes the system. 
The following choice \cite{Yamada2011},
\begin{equation}
F(r)=M_0r^3, \quad v_e(r)=r^\alpha, \quad F(R)=M_0R^\frac{3}{\alpha+1}
\label{toymodel}
\end{equation}
imply an equilibrium state for the evolving collapse configuration, where the collapse of matter cloud from a regular initial state settles to a final static solution which is an object-like naked singularity. The equilibrium metric in the interior 
$R<R_b$ of the boundary $R_b$, has a central naked singularity, and is given by
\begin{equation}
ds^2_e =  -(1-M_0)\left(\frac{R}{R_b}\right)^{\frac{M_0}{1-M_0}}dt^2+
\frac{dR^2}{1-M_0}+R^2d\Omega^2 \; .
\end{equation}
This can be matched smoothly to a Schwarzschild spacetime, or with FLRW geometry in the exterior $R\geq R_b$.
We thus have a one-parameter family of static equilibrium solutions
parametrized by $M_0$. Each member of this family of solutions has a naked singularity at the center. The solutions are
obtained as the end state of dynamical collapse from regular
initial conditions with $F(r)=M_0r^3$, and choosing the evolution
function $v(r,t)$ such that it asymptotes to the required
$v_e(r) \propto r^2$ as $t\to\infty$.

This equilibrium metric, obtained as the final state of collapse, is singular at center with the energy density diverging at $r=0$, thus presenting a central
singularity created as the result of gravitational collapse from regular initial data respecting the energy conditions.  Here the outer matter shells do not fall into the singularity but halt at a finite radius, thus creating a static compact object.

The equilibrium configuration resulting as collapse final state 
has a singularity at center, and we check the conditions when it is 
not covered by an event horizon. This means that the boundary where the cloud matches the exterior has a radius
greater than the Schwarzschild radius. Then the
boundary condition at equilibrium implies that for
the absence of trapped surfaces we must have
$F/R<1$, which is easily verified here.

We thus have the interesting result that
any central singularity that forms in the final equilibrium
state via this collapse dynamics is always a timelike naked singularity,  
and during entire collapse evolution the center is never trapped.

As we noted, the ultra-high density region that develops at the 
center of collapse continues to be always 
visible and never trapped. The classical relativity may eventually 
break down here at high enough densities and quantum gravity effects may
then dominate which would be visible. This phase is always obtained 
in a finite but large enough coordinate time before the actual singularity of the equilibrium. In contrast, in the case of collapse to a black hole,
the ultra-high density regions are always necessarily hidden
inside the event horizon after a certain stage of collapse, and
any Planck scale physics that might occur close to the singularity
is invisible to distant observers.

\subsection{Scalar Field Collapse}\label{Scalar}

Due to pre-dominance of various scalar fields in the early universe, a consideration of scalar field collapse scenarios is important towards the formation of PBHs/PNaSs. Scalar field collapse with non-zero potentials has been studied in recent years for a variety of cases, and the results show that naked singularities, either event-like or object-like, form in such a collapse 
in a set of non-zero measure \cite{Dey2023,Mosani2023}. 
There are classes of potentials that lead to naked singularities, just as there are classes that lead to black holes, each developing from physical initial conditions.
More detailed discussion of this will be taken up elsewhere.

Again, the collapse considered here is within an overall FLRW cosmological background geometry. Collapse models involving massless scalar fields including those with non-zero potentials and their connection to scalar field dark matter and dark energy is worth examination. This will be discussed in further detail elsewhere.

\section{Discussion}\label{Discussion}

We pointed out here that the primordial black holes are not the only consequence of gravitational collapse that occurs in the early universe due to density fluctuations, and that primordial naked singularities would form in many situations. Since gravitational collapse in general relativity does not always necessarily give rise to black hole as collapse final state, the occurrence of PBH or PNaS in the early universe and their abundances depend on the initial conditions within the collapsing density fluctuations. In fact, PNaS could occur in many physically realistic situations, as we find that while homogeneous initial density profiles give rise to black holes, a profile with higher central density leads to a naked singularity. 
Moreover, if early universe is dominated by a major component which is 
essentially pressureless non-interacting dark matter 
\cite{Yamada2011}, the collapse outcomes for dust models 
\cite{JoshiDwivedi1999,JoshiDwivedi1993,GoswamiJoshi2007} 
would need careful attention. The dust-like phase of early 
universe has been discussed (see, e.g. \cite{Harko2012} and 
references therein), which as we discussed would favour generation 
of PNaSs through gravitational collapse. 

Therefore, it is important to explore the astrophysical and cosmological 
implications of primordial naked singularities. Two main issues here 
are dark energy and dark matter. Their sources are completely unknown 
presently even though it is widely thought that both of them exist. 
Interestingly, in recent days there have been works that explore 
the particle creation near naked singularities in some detail 
(see, e.g., \cite{Henheiketal2024}).
It is seen that well-defined, self-adjoint Hamiltonians exist 
with particle creation at the naked singularity, showing that 
a naked space-time singularity need not lead to a breakdown 
of physical laws, but in fact allows for boundary conditions 
governing what comes out of the singularity, thereby removing 
the ultraviolet divergence. Such a direction may help us build 
phenomenological models for dark matter creation in the vicinity 
of a naked singularity. 
When event-like naked 
singularities arise, a visible curvature region develops with 
extreme values of densities, pressures, and other physical 
quantities. Copious particle creation may happen in such strong 
gravity regions with many quantum gravity dominated effects. 
The emitted radiations would have very high but finite red-shifts. 
Then possibly the PNaSs, or particle creation near the same could 
contribute to part of dark matter today. 
However, note that we do not exclude PBH or any other previously 
suggested constituents of the dark matter in this context, rather 
we propose that PNaSs are a viable dark matter candidate and could 
coexist with some other dark matter candidates. 
 
We now briefly discuss some implications of PNaSs in contrast to those of PBHs. We consider object-like PNaSs which are expected to exist till the current time.
Note that PBHs and PNaSs can have different mass spectra and abundance due to different formation conditions and evolution.
Varieties of constraints have been suggested for PBHs (see \cite{Carr2021} and references therein). Some of these should be similar for PNaSs, and some others
can be different. Note that while different constraints can distinguish between PBH and PNaS populations, the expected similar constraints imply that the already suggested constraints for PBHs should also somewhat apply to PNaSs.

PBHs could be detected or inferred through gravitational effects and radiation.
The radiation could be due to accretion of baryons over cosmic time and Hawking radiation, both can add to the observable radiation and particle background
(\cite{Carr2020,Hasinger2020} and references therein). Here, a gravitational constraint could be due to lensing, expected to be similar for PBHs and PNaSs of the same mass, unless the normal distance of the path of the wave (electromagnetic or gravitational) from the object and the wavelength are sufficiently small.
There could also be dynamical constraints (see \cite{Harko2012} and references therein) which can come from a wide variety of phenomena, such as (i) the
effects of passing PBHs/PNaSs by or through cosmic objects of
various types (e.g., stars, globular clusters) and (ii)
the capture of a PBH/PNaS by a star and a plausible transmutation of the star into a collapsed object. These constraints could be similar for PBHs and PNaSs for some cases.
However, while for a captured PBH, the stellar host may transmute into a black hole, the end product might be different for a captured PNaS.
Another constraint could come from the  mergers of collapsed objects \cite{Carr2020}. Note that PNaS--PNaS and PBH--PNaS mergers can be different from PBH--PBH mergers in terms of the observed properties of gravitational waves and electromagnetic waves. Particularly, while a PBH--PBH merger may not generate electromagnetic radiation, PNaS--PNaS and PBH--PNaS mergers could.
Moreover, accretion onto individual stellar-mass or larger collapsed objects could also distinguish a black hole from a naked singularity \cite{Chakraborty2017,Kocherlakota2019}.

Considering the radiation due to accretion of baryons over cosmic time, the observable background could distinguish between PBH and PNaS populations, if the accretion processes onto these objects are well understood. Moreover, one does not expect PBHs with mass $\lesssim 5\times10^{14}$~g to exist today due to Hawking radiation (but see \cite{Chakraborty2022}). However, as mentioned in 
section~\ref{Object}, 
the object-like PNaSs with mass $\lesssim 5\times10^{14}$~g could exist today, and could significantly contribute to dark matter. This and other constraints mentioned above imply that the maximum possible contribution to dark matter could be different for PBHs and PNaSs for a given mass range, which may have a significant impact on 
the dark matter physics.

Finally, we point out that naked singularities, in principle, can allow us
to access the region with extreme space-time curvatures and the
matter with extreme densities, which are not accessible due to black hole
event horizon. This could be useful to discover new physics and probe poorly 
understood aspects of the fundamental physics, such as quantum gravity.
A large number of existing PNaSs could be particularly helpful to achieve 
these scientific goals.
Furthermore, the idea of PNaSs might allow the tantalizing possibility that
a significant fraction of matter in the universe is in the form of 
visible singularities.

\acknowledgments

SB acknowledges the support of the Department of Atomic Energy (DAE), India.

\end{document}